\def \half {{1\over 2}}

\def\Square{{\vbox {\hrule height 0.6pt\hbox{\vrule width 0.6pt\hskip 3pt
        \vbox{\vskip 6pt}\hskip 3pt \vrule width 0.6pt}\hrule height 0.6pt}}}

\def\roughly#1{\mathrel{\raise.3ex\hbox{$#1$\kern-.75em
\lower1ex\hbox{$\sim$}}}}

\def\pref#1{(\ref{#1})}

\def\cM{{\cal M}}

\def\hf{{1\over 2}}
\def\Tr{{\rm Tr}}
\def\Dsl{\hbox{/\kern-.6700em\it D}} 
\def\dsl{\hbox{/\kern-.5300em$\partial$}}
\def\psl{\hbox{/\kern-.5300em$p$}}
\def\qsl{\hbox{/\kern-.5300em$q$}}
\def\sl#1{\hbox{/\kern-.5300em$#1$}}

\def\psibar{\overline{\psi}}

\def\eqa{\begin{eqnarray}}
\def\eeqa{\end{eqnarray}}
\def\eq{\begin{equation}}
\def\eeq{\end{equation}}

\def\nth#1{{1 \over #1}}
\def\endignore{}
\def\ignore#1\endignore{}
\def\det#1{|\!|#1|\!|}

\def\bfB{{\bf B}}
\def\bfE{{\bf E}}

\def\bfJ{{\bf J}}
\def\bfa{{\bf a}}

\def\bfe{{\bf e}}
\def\bfj{{\bf j}}

\def\bfn{{\bf n}}
\def\bfp{{\bf p}}
\def\bfq{{\bf q}}
\def\bfr{{\bf r}}

\def\z{\zeta}

\def\myref#1{(\ref{#1})}

\documentstyle[prl,aps]{revtex}

\begin{document}
\twocolumn[\hsize\textwidth\columnwidth\hsize\csname@twocolumnfalse%
\endcsname


\draft
\title{Effective Actions, Boundaries and Precision Calculations of Casimir Energies}

\author{Y. Aghababaie and C.P.~Burgess}

\address{Physics Department, McGill University,
3600 University Street, Montr\'eal, Qu\'ebec, Canada H3A 2T8.}
\maketitle

\begin{abstract}
{We perform the matching required to compute the leading effective
boundary contribution to the QED lagrangian in the presence of a
conducting surface, once the electron is integrated out. Our
result resolves a confusion in the literature concerning the
interpretation of the leading such correction to the Casimir
energy. It also provides a useful theoretical laboratory for
brane-world calculations in which kinetic terms are generated on
the brane, since a lot is known about QED near boundaries.}
\end{abstract}
\pacs{PACS numbers: 12.20.Ds, 03.70.+k \hfil Preprint McGill-03/05
\qquad arXiv:quant-ph/0304066}
%
%
]

\section{Introduction}
More than 50 years ago Casimir \cite{Casimir} forced physicists to
recognize the reality of quantum vacuum fluctuations by showing
that those of the electromagnetic field can mediate physical
effects, such as causing a force between two parallel plates. For
instance, for plane parallel metallic plates separated by a
distance $a$ the energy per unit area associated with this force
is\footnote{We use units for which $\hbar = c = 1$.}
\eq\label{E0}
    \varepsilon_0 = {E_0 \over A} = - \, {\pi^2 \over 720 \, a^3} \, .
\eeq

The study of this effect has experienced a recent revival,
largely due to prospects for its improved measurement. (For a
modern review see, for example, ref. \cite{BordagReview}.) With
this recent attention has come more detailed calculations,
including the one-loop corrections within Quantum Electrodynamics
(QED) due to virtual photons and electrons \cite{BRW}. For $ma \gg
1$ the leading correction found in this way is
\eq \label{E1}
    \varepsilon_1 = {\pi^2 \, \alpha \over 2560 \, m \, a^4} \, ,
\eeq
where $m$ is the electron mass and $\alpha = e^2/4\pi$ is the
usual fine-structure constant.

Surprisingly, a whiff of controversy has lingered over the
physical interpretation of eq.~\pref{E1}, a controversy which can
be traced to its dependence on the electron mass. The
controversial issue has been cast most sharply when the
calculation is formulated within an effective-lagrangian
framework, as might be expected to be appropriate
given that the electron mass is much higher than the energies, $E
\sim 1/a$, of the photon modes whose contributions dominate in the
Casimir effect. Early workers \cite{KongFinn} found a much smaller
contribution than eq.~\pref{E1}, with the leading effect instead
found to arise from the Euler-Heisenberg effective interaction
obtained by integrating out the electron, giving:
\eq \label{E1'}
    \varepsilon_1' = {11 \, \pi^4 \, \alpha^2 \over 3,888,000 \,
    m^4 \, a^7} \, .
\eeq

Subsequent workers have verified the calculation leading to the
result \pref{E1}, and have attributed the discrepancy either to a
failure of effective field theory itself \cite{BS} or to a
misidentification of the most important effective interaction
which is relevant \cite{Finn01,Melnikov}. There remains a
disagreement about which effective interaction is most relevant,
with ref.~\cite{Finn01} arguing that eq.~\pref{E1} can be
reproduced by an interaction localized on the plates (the
`boundary'), with an effective coupling which is of order
$\alpha/m$. By contrast ref.~\cite{Melnikov} argues that the
required effective interaction arises in the space between the
plates (the `bulk'), with a coupling which is of order $\alpha /(m
\, a)$. Unfortunately, neither reference resolved the discrepancy
by performing the matching calculation which is required in order
to properly identify which effective interactions actually arise
in the low-energy theory. (See refs.~\cite{Andersen,CS} for other
discussions of effective lagrangians in the Casimir energy
problem.)

It is our purpose with this paper to settle the issue of which
electron-mass corrections are dominant, and how they arise within
an explicitly-constructed effective field theory. In order to do
so, we set up the relevant effective lagrangian, and perform the
requisite matching calculation which determines the size of the
effective couplings. We draw the following conclusions:
\begin{itemize}
\item We conclude that the correct effective operator is the local
boundary operator of ref.~\cite{Finn01}, which describes how the
vacuum polarization alters the interaction between a test charge
and the surface charges which it induces on the conductor.
\item As the matching calculation shows explicitly, this effect
relies on polarizations of the vacuum charge over distances of
order $1/m$, and so presupposes that the conductor's boundaries
are sharp on these scales. As such they are likely to be dominated
by surface effects for real conductors, for which the scale of
penetration depths for electromagnetic fields are set by much
larger interatomic separations, of order $1/(\alpha \, m)$.
\end{itemize}

We present our results in more detail in the next sections,
starting with a brief summary of some general features which all
effective field theories must share. We then identify the dominant
effective operator for the Casimir effect which arises when the
electron is integrated out. Finally we perform the matching
calculations which are required to identify the dominant electron
contributions to the Casimir energy in both of these cases. Our
conclusions are briefly summarized at the end.

\section{The Effective Field Theory}
Although their use has been largely systematized only during the
past 20 years, the main ideas of effective field theories go back
to the much earlier Born-Oppenheimer approximation used in atomic
physics decades earlier. The idea is to exploit the way physical
systems simplify when they are probed only at very low energies,
$E$, compared with some intrinsic energy scale, $M$. Expressions
for general low-energy observables usually greatly
simplify once they are expanded in powers of $E/M$, and so it pays
to take advantage of this expansion as early as possible in a
calculation.

The contribution of field theory to this process comes with the
recognition that all of the low-energy effects of virtual
high-energy states, $h$, on low-energy degrees of freedom, $\ell$,
can always be expressed in terms of local operators involving only
the light states $\ell$. That is, the physics of the full
hamiltonian
\eq \label{LowEH}
    H_{\rm full}(\ell,h) = H_{\rm l.e.}(\ell) + H_{\rm h.e.}(h)
    + H_{\rm int}(\ell,h) \, ,
\eeq
is indistinguishable from the physics of the effective hamiltonian
\eq
    H_{\rm eff}(\ell) = H_{\rm l.e.}(\ell) +
    \sum_{k\ge k_0} {1 \over M^k} \sum_l
    c_{kl} {\cal O}_{kl}(\ell) \, ,
\eeq
to {\it any fixed order} in powers of $1/M$, for some choice for
the effective interactions, ${\cal O}_{kl}$, and effective
couplings, $c_{kl}$. Furthermore, since the uncertainty principle
only permits energy and momentum conservation to fail over very
short times and short distances, the effective interactions,
${\cal O}_{kl}$, which are required are all {\it local} in space
and time --- {\it i.e.} involve products of fields and their
derivatives at fixed positions in space and instants of time.

The utility of this observation is that the required effective
interactions and couplings can be computed once and for all by
comparing to simple observables, and once obtained may be used for
{\it any} calculations in the low-energy theory. Such a
determination of the effective interactions and couplings is
called a `matching' calculation, because the effective
interactions of $H_{\rm eff}$ are matched onto what is required by
the full microscopic theory, $H_{\rm full}$. (See
refs.~\cite{EFTReviews} for reviews on effective field theories.)

\subsection{Effective Field Theories and QED}
Quantum electrodynamics in particular lends itself to this kind of
effective analysis, because of the huge hierarchy in scales
between the electron mass, $m$, and the other scales of usual
interest such as those appropriate to the propagation of light or
to atomic energy levels. Two kinds of effective field theories
have been explicitly treated in this way. One corresponds to
integrating out electrons and positrons and high-energy photons to
describe the interactions of low-energy photons \cite{LowEQED},
and the other involves integrating out positrons and high-energy
electrons and photons to describe the low-energy interactions of
nonrelativistic electrons and low-energy photons \cite{NRQED}.

For electromagnetism it is more convenient to work with an
effective action or lagrangian than with an effective hamiltonian.
In the absence of all boundaries and charge distributions,
integrating out the electron leads to the following Maxwell plus
Euler-Heisenberg \cite{LowEQED} effective interactions for
low-energy photons
\eqa \label{EHLag}
    {\cal L}_{\rm eff} &=&  \nth2 \, \left( \bfE^2 - \bfB^2
    \right)\cr
    && \qquad + {2 \alpha^2 \over 45 m^4} \, \left[
    \left( \bfE^2 - \bfB^2 \right)^2 + 7 \, \left( \bfE \cdot \bfB
    \right)^2 \right] + \cdots
\eeqa
In this expression, corrections to the first (Maxwell) term in
this lagrangian have been removed using an appropriate rescaling
of the electromagnetic field. Similarly, a possible
$O(\alpha/m^2)$ (Uehling) term, of the form $F^{\mu\nu} \Square \,
F_{\mu\nu}$ arising from the vacuum polarization, is not written
here because it can be removed to this order by performing a field
redefinition of the form $A_\mu \to A_\mu + {c_1 \alpha \over m^2}
\, \Square \, A_\mu + {c_2 \alpha^2 \over m^4} \, \Square^2 \,
A_\mu$, for an appropriate choice of coefficients, $c_1$ and
$c_2$. Any effective interaction which can be removed in this way
is called `redundant', since it cannot contribute to physical
quantities.

The numerical coefficient of the quartic (Euler-Heisenberg) term
may be found by comparing the amplitude for light-by-light
scattering as computed to order $\alpha^2/m^4$ using eq.~\pref{EHLag}
and with the full QED lagrangian, including electrons.
Once this is done, the result captures the influence, to order
$\alpha^2/m^4$, of electrons on {\it any} low-energy observable
(in the absence of boundaries) because
eq.~\pref{EHLag} completely exhausts the possible local,
Lorentz-invariant and gauge invariant effective interactions which
may be constructed to this order.

\subsection{Boundary Charges and Screening}
For applications to the Casimir effect we must ask how the
presence of the conducting boundaries can affect the matching
process just described. For simplicity we examine the case
considered by refs.~\cite{BRW,KongFinn,BS,Finn01,Melnikov} and
restrict our discussion to the case where the electrons do not
`see' the boundaries, and only the boundary conditions of the
electromagnetic field are changed. For conducting plates we take
the electromagnetic boundary conditions to be
\eq \label{CondBC}
    \left. \epsilon^{\mu\nu\lambda\rho} n_\nu F_{\lambda\rho}
    \right|_{\partial \cM} = 0 \, ,
\eeq
where $n_\nu = \{0,\bfn\}$ is a normal vector on the surface which
points into the bulk. This captures the usual conducting boundary
conditions, that $B_n = \bfn \cdot \bfB$ and $\bfn \times \bfE$
both vanish. $\partial\cM$ here denotes the surfaces of the
plates, considered as the boundaries of the intervening bulk,
$\cM$.

Our starting point is then the QED action, with boundary terms
\eqa \label{Action}
    S_{\rm full} &=& -  \int_{\cM} d^4x \; \left( \nth4 F^{\mu\nu}
    F_{\mu\nu} + \psibar \, (\Dsl + m) \psi \right) \cr
    && \qquad - \int_\cM d^4x \; J^\mu \, A_\mu -
    \int_{\partial{\cal M}} d^3x \; j^\mu \, A_\mu \, ,
\eeqa
where $D_\mu  = \partial_\mu  + i e A_\mu$ is the usual covariant
derivative for the electron field. Here $J^\mu = \{\rho,\bfJ\}$
denotes any classical test charges and currents with which we
choose to probe the system, and $j^\mu = \{\sigma, \bfj \}$
denotes the surface charge and current densities whose presence
enforces the boundary condition, eq.~\pref{CondBC}. The classical
Maxwell equations obtained by varying eq.~\pref{Action} with
respect to $A_\mu$ in the bulk {\it and} on the boundaries is:
\eqa \label{ClassEQ}
    \partial_\mu F^{\mu \nu} - J^\nu = 0 \qquad &&\hbox{in ${\cal M}$;}
    \cr
    n_\mu F^{\mu\nu} - j^\nu =0 \qquad &&\hbox{on $\partial{\cal
    M}$.}
\eeqa
The boundary part of these equations relate the surface charges
and currents to the boundary electric and magnetic fields and are
more familiar once written explicitly in terms of $\bfE$ and
$\bfB$. For the surface charge distribution they imply, for
instance,
\eq \label{Ensigma}
    \left. \bfn \cdot \bfE \right|_{\partial\cM} = \left. E_n
    \right|_{\partial \cM} = \sigma \, .
\eeq

The presence of these boundary charges plays a crucial role in
constructing the effective theory, because of its interplay with
the photon's vacuum polarization. To see this imagine we now
compute the electron's contribution to the vacuum polarization
\eqa
    &&\Pi_{\mu\nu}(q^2) = \left(q^2 \eta_{\mu\nu} -
    q_\mu q_\nu\right) \Pi(q^2) \cr
    && \quad = -i e^2 \int {d^4p  \over (2\pi)^4}
     \, \Tr{\left[ {-i\psl +m \over p^2 + m^2_\epsilon}\,
    \gamma_\mu\, {-i(\psl - \qsl) + m \over (p-q)^2 + m^2_\epsilon}
    \, \gamma_\nu \right]}
\eeqa
with $m^2_\epsilon = m^2 - i \epsilon$. After renormalization,
$\Pi(q^2)$ is given by
\eq
    \Pi(q^2) = {2 \alpha \over \pi} \int_0^1 \, d\z \, \z(1-\z) \,
    \log \left[1 + \z(1-\z) {q^2 \over m^2} \right].
\eeq

$\Pi(q^2)$ describes the polarization of the vacuum about any given
charge distribution, effectively smearing it over a distance of
order $1/m$. Given a point charge source, $Q \, \delta^3(\bfr)$,
the position-space charge density which the electron vacuum
polarization produces is (to leading order) $\rho_{\rm eff} = Q \,
\left[ \delta^3(\bfr) + \eta(\bfr) \right]$, where \cite{Wbg}
\eqa \label{eq:eta}
    &&\eta(\bfr) = {1\over 2} \int {d^3q \over (2\pi)^3} \; e^{i{\bf q}\cdot
    {\bfr}} \, \Pi(\bfq^2) \cr
    &&= {\alpha \over \pi } \int_0^1 d\z\, \z(1-\z)\, \cr
    && \qquad\qquad \times \int{d^3q \over
    (2\pi)^3} \; e^{i{\bf q}\cdot {\bfr}} \,
    \,\log\left[ 1+ {\bfq^2 \over m^2} \, \z(1-\z) \right] \\
    &&= N \, \delta^3(\bfr) - \, {\alpha \over 2 \pi^2 \, r^3}
    \int_0^1 \, d\z \; \left( 1 + {mr \over \sqrt{\z (1 - \z)}}
    \right) \cr
    &&\qquad\qquad \qquad\qquad\times\,  \z (1 - \z) \,
    \exp\left[ - \, {mr \over \sqrt{\z (1 - \z)}} \right] . \nonumber
\eeqa
Here $r = |\bfr|$ and $N$ is a constant which renormalizes the
bare charge $Q$, determined by the condition
\eq
    \int d^3r \, \eta(\bfr) = \half \, \Pi(\bfq^2 = 0) = 0.
\eeq

For our purposes what is important is that virtual electrons also
act to screen the surface charges which are required at the
boundary $\partial\cM$ to enforce conducting boundary conditions
there. The resulting charge distribution may be obtained by
integrating eq.~\pref{eq:eta} over a planar sheet of charge,
$\sigma \, \delta_+(z)$, where $\delta_+(z)$ is normalized so it
integrates to unity on one side of the boundary: $\int_0^\infty dz
\; \delta_+(z) = 1$. (For instance, this can be represented by
$\delta_+(z) = \lim_{\lambda \to \infty} \left(\lambda \,
e^{-\lambda |z|}\right) = 2 \, \delta(z)$.) We write, then
\eq
    \sigma \, \delta_+(z) = 2 \sigma \delta(z) = 2 \sigma \int
    d^2a
    \, \delta^3(\bfr - \bfa) \, ,
\eeq
where we take $\bfa$ to be a vector lying on the conducting
surface and $z$ to be the coordinate in the direction
perpendicular to this surface. Using this with eq.~\pref{eq:eta}
gives the result for the resulting polarized charge distribution
around a surface-charge sheet positioned at $z=0$
\eqa \label{Poldist}
    \rho(z) &=& \sigma \left\{ (1+N) \, \delta_+(z)
     + {2 \, \alpha \over \pi}
    \int_0^1 d\z \,\z(1-\z) \right. \cr
    &&\qquad \left.\times  \int {dq_z \over 2\pi} \, e^{iq_z z}
    \log\left[1+\z(1-\z) {q_z^2 \over m^2}\right] \right\} \\
    &=& \sigma \left\{ (1 + N) \, \delta_+(z)\phantom{\frac12} \right. \cr
    && \qquad \left. - {2 \, \alpha \over \pi \, |z|}\,
     \int_0^1 d\z \, \z(1-\z)\, \exp\left[ - \, {m |z| \over
    \sqrt{\z(1-\z)}}\right] \right\} . \nonumber
\eeqa

The first term in eq.~\pref{Poldist}, involving the delta
function, expresses how virtual electrons renormalize the bare
surface charge distribution. Of more interest for the present
purposes is the second, position-dependent component. 
This polarization charge distribution
has the physical effect of generating multipole moments around the
uniform surface charge at $z=0$, which are detectable (in
principle) through their interactions with probe charges in the
bulk. In the effective theory obtained when the electron is
integrated out, the effect of these multipole moments must be
replaced by effective interactions which are localized on the
surface of the conducting plates.
%
%

The particular moment of interest in what follows is the electric
dipole moment density, $\bfp$, which this charge density defines
\eqa \label{Dresult1}
    \bfp(z) &=& \rho(z) \, z \, \bfe_z \\
    &=& - \, {2 \, \alpha \, \sigma \, \bfe_z \over \pi} \int_0^1 d\z \,
    \z(1-\z)\, \exp\left[ - \, {m |z| \over
    \sqrt{\z(1-\z)}}\right] \, , \nonumber
\eeqa
and which for large $m$ becomes
\eqa \label{Dresult}
    \bfp(z) &\to& - \, {2 \, \alpha \, \sigma \, \bfe_z \over \pi  m}
    \, \delta_+(z) \int_0^1 d\z \, \Bigl[\z (1-\z) \Bigr]^{3/2}
    \cr
    &=& - \, {3\, \alpha \, \sigma \, \bfe_z \over 64 \, m} \, \delta_+(z) \, .
\eeqa

This polarization distribution introduces a surface contribution
to the field energy, $U$, when a test charge, $Q$, is placed in
the vicinity of a conducting plate. In the absence of vacuum
polarization we have seen that any test charge induces a nonzero
charge density, $\sigma$, on the conductor's surface, as given by
eq.~\pref{Ensigma}. Virtual electrons then polarize the vacuum
within a distance $1/m$ of both the test charge {\it and} this
surface charge, leading to a change in the field energy. To
leading order this change is the sum of the interaction of the
induced charge density $\sigma$ with the polarization around the
test charge, plus the interaction of the test charge $Q$ with the
induced polarization near the surface charge. Each of these
effects has precisely the same size,\footnote{The equality of the
two contributions is most easily seen if the surface charge is
instead represented as an equivalent image charge, on the opposite side of
the boundary.} leading to a correction of
the interaction energy given by
\eqa \label{Echange}
    \Delta U &=& 2 \times \left(\frac12\right)
    \int d^3r \; \bfE \cdot \bfp \cr
    &=& - \, {3\, \alpha \over 64 \, m} \int d^2r \; \sigma \, E_n
    \cr
    &=& - \, {3\, \alpha \over 64 \, m} \int d^2r \; E_n^2 \, ,
\eeqa
where we use the lowest-order result, eq.~\pref{Ensigma}, to write
$\sigma = E_n$. Here $\bfE$ is the lowest-order electric field
(or electric displacement) not including the vacuum-polarization corrections.

\subsection{Matching Conditions}
We now consider integrating out the electron, and ask what
effective interaction in a low-energy theory without electrons
describes their effects to leading order in $1/m$. As we have seen
above, in the absence of the charge densities on the surfaces of
the conducting plates, in the bulk the vacuum polarization
provides only interactions of the form $F^{\mu\nu} F_{\mu\nu}$ or
$F^{\mu\nu} \Square \, F_{\mu\nu}$, which are redundant
interactions with no physical consequences. The vacuum
polarization {\it does} give nontrivial contributions once the
surface charges are considered, however. As discussed above, it
generates multipole moments along all of the conducting surfaces
due to the induced bulk charge redistribution it implies.

For plane parallel conducting plates the symmetries of the problem
require that the operator obtained must be translation- and
Lorentz-invariant within the dimensions parallel to the plates, as
well as being parity and time-reversal invariant. Keeping in mind
that $\bfE_\parallel$ and $B_n$ both vanish on the conductors, the
lowest-dimension operator which is possible involving the
electromagnetic field is
\eq \label{BoundOp}
    \Delta S = \half \int_{\partial\cM} d^3x \; \left[ c_1 \, E_n^2 +
    c_2 \, \bfB^2_\parallel \right] \, ,
\eeq
where $c_1$ and $c_2$ are constants to be determined. Since the
electrons do not themselves see the boundary, their lowest-order
contribution is actually Lorentz-invariant in all 4 dimensions,
allowing the simplification $c_2 = - c_1$.

The coefficient $c_1$ may be determined by computing the
contribution this operator makes to the field energy density,
$u({\bf r})$, of a classical static test charge, giving
\eq
    \Delta u = {c_1 \over 2} \; E_n^2 \,
\eeq
which when compared with eq.~\pref{Echange} (using $U = \int d^2r
\; u$) gives the result
\eq \label{c1val}
    c_1 = -\, {3 \alpha \over 32 \, m} \, .
\eeq

We see that the coefficient of this operator may be obtained for
any conductor independent of the presence of any other conductors,
because it describes a local condition --- the vacuum polarization
--- near the conductor's surface. Once it is determined, its
physical origin may be forgotten and its influence in any other
low-energy process may be obtained perturbatively in the
coefficient $c_1$.   In particular, the effective operator with the
coefficient $c_1$ given in \pref{c1val} may be used to calculate
the leading correction to the Casimir energy.  This has been
performed in ref.~\cite{Finn01} and agrees with the calculation
in the full theory~\cite{BRW}, as it must.

\section{Conclusions}
The general utility of effective field theories lies in their
efficient identification of the scales which are relevant to any
particular physical system, and we see that this is also true for
the Casimir energy. The novel feature which conducting boundaries
introduce into the low-energy QED effective theory is the presence
of surface charges and currents, whose presence provides a way for
the vacuum polarization to have physical implications which it
would not otherwise have. Its effects are described by local
boundary interactions because the vacuum polarization only extends
over distances of order $1/m$, it so cannot reach very far into
the bulk. Of course, it should be emphasized that for real systems
the effective interactions computed in this article are not the
most important, since other microscopic physical length scales
arise which are much larger than the electron's Compton
wavelength.

Because effective field theories are so easy to use, they normally
really come into their own once one proceeds beyond leading order
in small quantities like $\alpha$ or $1/m$, since they make
possible calculations which would otherwise be impractical. This
is likely also to be true for precision calculations of Casimir
energies, for which there is now considerable practical interest
in understanding the dependence of the effect on the geometry and
physical make-up of the conductors and dielectrics involved.

For real systems there are a number of effects whose contributions
to the Casimir energy must be disentangled, and since each comes
with an associated length scale which is short compared with the
inter-plate separation it is likely most efficient to do so within
an effective-field-theory analysis. These effects include thermal
fluctuations; shape effects due to the curvature or roughness of
the conducting surface; and effects due to imperfections in the
ideal-conductor boundary conditions. Conductivity effects can be
associated with the fact that static external electric fields have
a finite penetration length into the conductor, and that
conduction electrons cannot adjust quickly enough to respond
perfectly to fields which oscillate sufficiently
rapidly.\footnote{For recent calculations discussing loop
contributions including boundary effects, see \cite{jaffe} and
references therein.}. Each of these should correct the Casimir
energy by amounts proportional to powers of small ratios of the
form $\lambda/a$, where $\lambda$ is the length scale relevant to
the microscopic physics of interest, and their dominant effects
can be parameterized in terms of local $\lambda$-dependent
effective interactions.

In general the dominant corrections must correspond to the
lowest-dimension effective interactions, some of which we now
display. Effective interactions which describe some conductivity
effects take a form similar to those considered here,
eq~\myref{BoundOp},
\eq
    \Delta S = -\hf\int_{\partial M} d^2r \left[ c_1 E_n^2
    + c_2 {\mathbf{B}}_{\parallel}^2 + c_3 {\mathbf E}_{\parallel}^2 +
    c_4 B_n^2 \right].
\eeq
On dimensional grounds one expects the coefficients of these
operators to receive effects of order $c_i \sim {\lambda}$. Notice
that the fields $\mathbf{E}_{\parallel}$ and $B_n$ can now appear
here because in real systems the conducting boundary conditions
are only imperfectly imposed.

Geometrical effects associated with the curvature of the
conducting surface may be similarly parameterized in terms of a
general effective lagrangian built from the surface's intrinsic
and extrinsic curvature tensors, ${R^a}_{bcd}$ and $K_{ab}$.
Sample low-dimension terms include
\eq
    \Delta S = \int_{\partial M} d^2r \; \sqrt{\det g}
    \left[s_0 + s_1 K + s_2 R +
    \cdots \right]\,,
\eeq
where $R = g^{bd}{R^a}_{bad}$, $K = g^{bd}K_{bd}$ and ${\det g} =
\hbox{det} (g_{bd})$, for $g_{bd}$ the induced metric on the
surface. On dimensional grounds one also expects the effective
couplings, $s_k$, to be proportional to powers of the relevant
microscopic length scale, $\lambda$.

Besides its practical applications, the Casimir energy system also
provides a useful theoretical framework in which to test some of
the more speculative theoretical ideas which have gained currency
of late. Similar issues to those considered here arise in
brane-world scenarios, wherein ordinary particles are confined to
surfaces (branes), with various interactions probing the bulk.
Renormalization issues can also have practical implications in
this case \cite{Ira}, and QED provides a useful benchmark against
which these more speculative calculations can be tested.

\section{Acknowledgements}

We would like to thank Finn Ravndal and Ira Rothstein for very
helpful conversations about how to apply effective field theories
to the Casimir energy. Our research is partially funded by NSERC
(Canada), FCAR (Qu\'ebec) and by McGill University.

\end{document}